# Reactive species involved in higher seeds germination and shoots vigor through direct plasma exposure and plasma-activated liquids

T. Dufour[1], S. Zhang[1], S. Simon[1], A. Rousseau[1]

[1]LPP, CNRS, UPMC Univ Paris 06, Ecole polytechnique, Univ. Paris-Sud, Observatoire de Paris, Université Paris-Saclay, Sorbonne Universités, PSL Research University, 4 place Jussieu, 75252 Paris, France
Corresponding author: thierry.dufour@upmc.fr

**Abstract:** Cold atmospheric plasma treatments have been applied on lenses seeds and shoots to improve their germination and vigor rates. Two approaches have been considered: direct plasma exposure and plasma activation of liquids (tap water, demineralized water and liquid fertilizer). A special focus has been drawn on reactive oxygen species generated in the plasma phase but also in plasma activated media to understand their impact on germination process as well as on plants growth.

**Keywords:** plasma agriculture, reactive oxygen species, germination, DBD, plasma jet

## 1. Introduction

Agriculture is a ten thousands years old practice whose main technical prowess has appeared the last centuries. Traditional agriculture has been dethroned by chemical agriculture which in turn has been competed by inorganic agriculture to get rid of pesticide residues and chemical fertilizers. Since chemical agriculture raises heavy ecological issues and since inorganic agriculture ensures only lower productivity, new solutions are expected. Plasma agriculture appears as an innovative approach to overcome recent agricultural issues of food safety and productivity. Several works have already evidenced how cold atmospheric plasma (CAP) treatments could present a great potential in promoting seeds germination, stimulating plants growth, tailoring plants dormancy or even improving plants resistance to common diseases. Wheat seeds have been treated by Dobrin et al. using a surface DBD [1]. A 15-minute plasma exposure yields to a root length distribution centred on 36.5 cm (instead of 32.8 cm without plasma treatment). With the same type of atmospheric process, M. Cernak's team has succeeded in increasing the germination rate but also the dry mass and the vigor of the seedlings for treatment times typically less than one minute [2]. In addition, populations of epiphytic bacteria, phytopathogenic and toxinogenic filamentous fungi have been inactivated, thus inhibiting the growth of surface biofilms in seeds.

In these research works, we have selected lenses as plant models and designed plasma processes to improve their seeds germination and shoots vigor. The objective is also to bridge plasma science to agronomy and understand the impact of plasma physico-chemical properties on physiological plants features, with a special focus on reactive oxygen species generated by the plasma phase.

## 2. Experimental setups

Two plasma sources have been designed and built to apply cold atmospheric plasma on seeds and shoots: a plasma jet supplied in helium at 200-500 Hz (AC) as shown in Fig. 1., and a planar dielectric barrier discharge operating on the same AC frequency range. Voltage, current and plasma power have been determined using 10 nF capacitors and Lissajous curve method. The treatments are carried out following two approaches: a direct approach where seeds are directly exposed to the plasma phase and an indirect approach where they are immerged into a liquid exposed to the plasma. Once germination has occurred, shoots are irrigated with plasma activated media (PAM). Three types of liquids have been tested: tap water (TW), demineralized water (DEM) and garden liquid fertilizer (FTZ). These liquids are considered as controls (CTRL) if not plasma-treated and qualified as plasma activated media (PAM) in the other way. During 2 weeks, shoots are daily (i) exposed 8h to artificial lamps dedicated to agronomic applications and (ii) irrigated by 5-10 mL of CTRLs or PAMs. Roots and stems lengths are measured as well as dried masses to attest the boosting plasma effect.

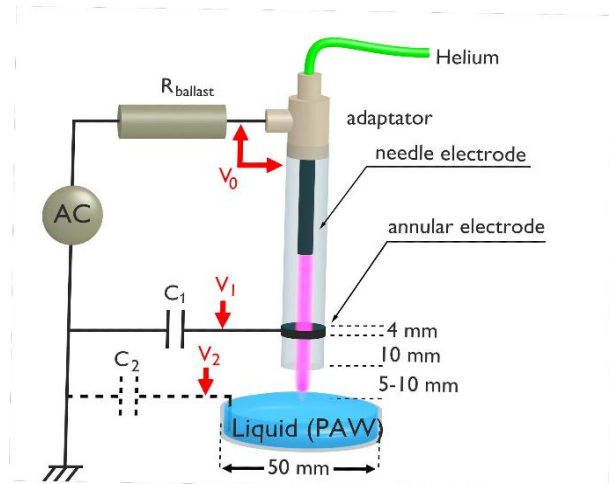

Fig. 1. Plasma jet setup for the activation of liquids dedicated to seeds and shoots.

Chemical species have been investigated in the plasma phase combining optical emission spectroscopy and mass spectrometry but also in the liquid phase using spectrophotometry techniques [3] and liquid probes so as to measure hydrogen peroxide, nitrates and nitrites. The PAMs have also been investigated using various chemical probes to measure pH and redox potentials.

## 3. Results & discussion

### 3.1. Plasma jet in spark-to-streamer mode

PAMs have been prepared with the jet device operating in a spark-to-streamer mode rather than the usual streamer mode. The power delivered to the plasma jet is the same in the two modes but the radical species generated are different – at least in terms of concentrations – owing to a different setup configuration: the spark-to-streamer mode requires the liquid to be grounded while it has to stay at a floating potential in the streamer mode. As an example, we have drawn in Fig. 2., the concentration of hydrogen peroxide in tap water (after its plasma activation) as a function of treatment time considering the two modes. For 10 min of treatment time, the amount of $H_2O_2$ is almost 8 times higher using the spark-to-streamer mode compared with the streamer mode. Also, the deposited powers remain quite low in both cases (<500 mW) while the energetic yield is 35 nmol/J in the spark-to-streamer mode versus only 7 nmol/J in the streamer mode.

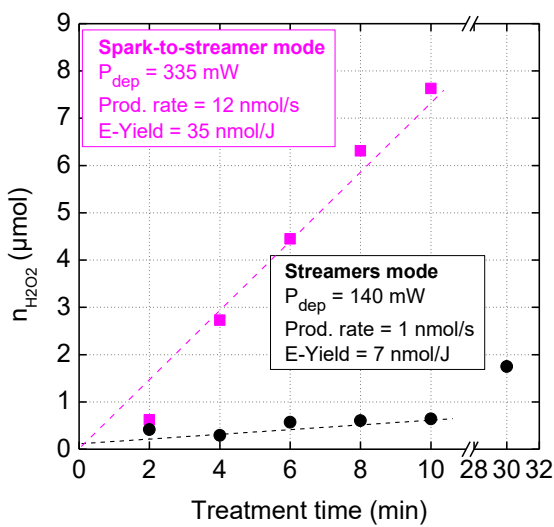

Fig. 2. Hydrogen peroxide produced in the liquid phase versus its plasma exposure time. The plasma jet is utilized either in spark-to-streamer mode or in streamer mode.

### 3.2. Plasma-assisted germination of lenses seeds

To study the effect of plasma on germination rate, lenses seeds are distributed in 6 test tubes (20 seeds/tube), as shown in Fig. 4. Each tube is filled with a specific liquid (10 mL), i.e. a control liquid or its corresponding PAM. No soil, no manure or any solid organic substrate has been utilized for sprouting and shoots growth.

The germination rate has been estimated at day 10 and is reported in Fig. 2. for the six previous experimental conditions. If the plasma treatment of demineralized water has no positive effect on the germination rate, it however has an important boosting impact with the two other PAMs. Indeed, the plasma activation of tap water has leaded to a germination rate increase from 22% to 100% and a similar trend has been obtained with the liquid fertilizer. Those rises result from complex mechanism between the native liquid oligo-minerals, the extern envelope properties of the seeds and the reactive oxygen species generated in the liquid phase upon its plasma exposure.

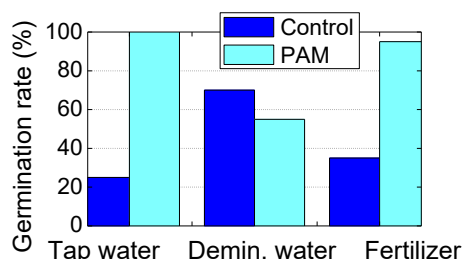

Fig. 3. Germination rates of lenses seeds considering tap water, demineralized water or liquid fertilizer either as controls or as plasma-activated media.

### 3.3. Influence of PAMs on lenses shoots stems

Then, shoots stems have been measured versus time, considering tap water, demineralized water and liquid fertilizer, with/without plasma exposure. The picture in Fig. 4. shows lenses shoots at day 12. As shown in Fig. 5, upon the 0-15 days observation window, tap water treatment is unsurprisingly less efficient than fertilizer. This is particularly true on the first days after germination but not at the end, e.g. stems are quite close in length at day 14. However, the plasma activation of these two liquids has a dramatic effect on lenses stems. For instance at day 14, stems are as high as 11.5 cm for the two PAMs, versus approx. 6.2 cm for the two controls. These trends have been confirmed by repeating three times the experiment.

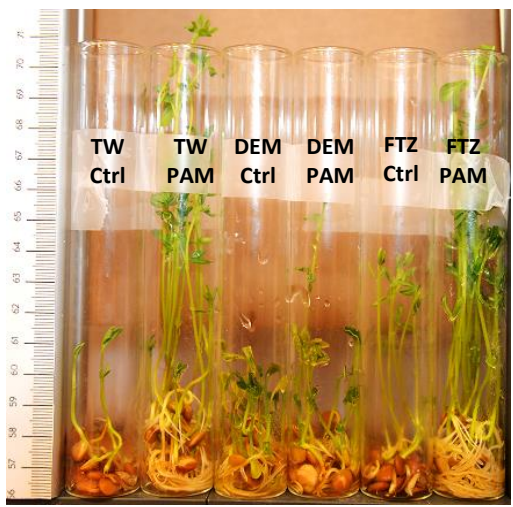

Fig. 4. Shoots at day 12. TW: tap water, DEM: demineralized water, FTZ: liquid fertilizer, Ctrl: Control, PAM: plasma-activated medium.

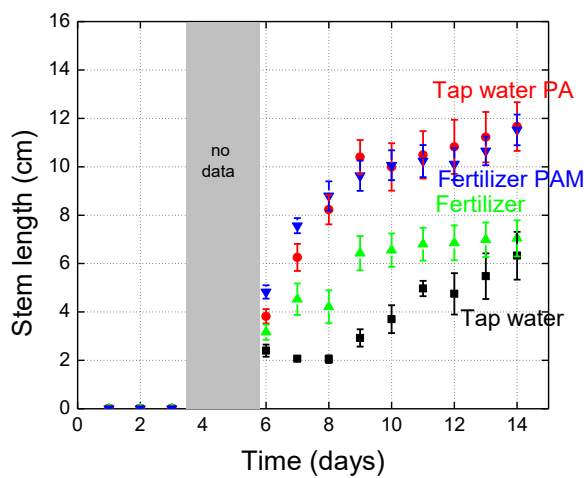

Fig. 5. Germinated lenses stems vs time. Shoots are daily irrigated with tap water/liquid fertilizer that are either untreated (controls) or treated (PAM).

To understand how the plasma activated media can stimulate both germination and shoots stems, hydrogen peroxide ($H_2O_2$), nitrites ($NO_2^-$) and nitrates ($NO_3^-$) have been measured in PAMs. $H_2O_2$ concentrations as high as 100 µM are evaluated in tap and demineralized waters using spectrophotometry (TiOSO$_4$ method [3]). This method could not been used to measure $H_2O_2$ in the liquid fertilizer owing to its initial coloration. Nitrite concentration as high as 100-150 µM have been estimated in plasma activated tap water and fertilizer vs approximately 50 µM in demineralized water. Finally, nitrates have been evaluated to very low concentrations in all cases (always <1µM) and do not appear as a species playing a crucial role in germination.

### 3.4. Forthcoming results: DBD in direct/indirect approaches

Since DBDs do not require a plasmagen gas (Ar, He) to operate, they may be considered as more economical than plasma jets. Their efficiency to produce reactive oxygen species will therefore be studied and compared with the E-yields and production rates obtained from the current plasma jet. Two approaches will be investigated: seeds directly exposed to the plasma phase and seeds immerged in liquids for a plasma activation.

Radicals such as OH, O and NO will be studied in the plasma phase by optical emission spectroscopy and mass spectrometry. Dedicated processes will be engineered to increase the density of a specific radical (and to the detriment of the others), to better understand its individual impact on seeds physiological parameters. A discussion will be also addressed to identify the reactive species appearing as promoters or inhibitors for germination and vigor. Finally, the issue of seeds selectivity to a same plasma treatment will be addressed by treating barley and sunflowers seeds.

### 4. Acknowledgements

This work has been done within the LABEX Plas@par project, and received financial state aid managed by the Agence Nationale de la Recherche, as part of the programme "Investissements d'avenir" under the reference ANR-11-IDEX-0004-02. This work has also been supported by the Île-de-France Region in the framework of the $^{PF2}$Abiomede Sesame project (16016309)